\documentclass{llncs}
\usepackage{graphicx}
\usepackage{latexsym}
\usepackage{textcomp}
\usepackage{graphicx}
\usepackage{latexsym}
\usepackage{textcomp}
\usepackage{wrapfig}
\usepackage{graphicx}
\usepackage{wrapfig}

\usepackage{latexsym}
\usepackage{textcomp}
\usepackage{listings}
\usepackage{color}
\usepackage{footnote}
\usepackage{multirow} 
\usepackage{array}
\usepackage{xspace}
\usepackage{units}

\usepackage[english]{babel}

\usepackage{amsmath, amssymb}

\usepackage{alltt}
\usepackage{algorithmic}
\usepackage{listings}

\usepackage{placeins}

\lstdefinelanguage{pseudo}{
morekeywords={if, else, for, while, remove, from, case, do, forever, to,
  transfer, function, return, in, then, elsif, endif, endwhile, endfor, forall, endforall},
sensitive=true,%
morecomment=[l]{//},%
morestring=[b]",%
}
\newcommand{\LT}{\textit{LT}\xspace}

\newcommand{\comment}[1]{ }

\newcommand{\suc}{\mathrm{suc}}
\newcommand{\sons}{\mathrm{sons}}
\newcommand{\dec}{\mathrm{dec}}
\newcommand{\inc}{\mathrm{inc}}
\newcommand{\set}{\mathrm{set}}
\newcommand{\father}{\mathrm{father}}
\newcommand{\KS}{\ensuremath{\mathit{KS}}\xspace}
\newcommand{\PKS}{\ensuremath{\mathit{PKS}}\xspace}

\title{An Experiment on Parallel Model Checking\\
  of a CTL Fragment\thanks{\footnotesize{This work was partially
      supported by JU Artemisia project CESAR, AESE project Topcased
      and R\'egion Midi-Pyr\'en\'ees}}} \titlerunning{}
\author{Rodrigo	T. Saad, Silvano Dal Zilio and Bernard Berthomieu \\
  \institute{CNRS, LAAS, 7 avenue du Colonel Roche, F-31400 Toulouse France \\
    Univ de Toulouse, LAAS, F-31400 Toulouse, France}
  \email{\{rsaad, dalzilio, Bernard.Berthomieu\}@laas.fr} } \titlerunning{}
\authorrunning{R. T. Saad, S. Dal Zilio and B. Berthomieu}

\begin{document}

\maketitle

\begin{abstract}
  We propose a parallel algorithm for local, on the fly, model
  checking of a fragment of CTL that is well-suited for modern,
  multi-core architectures. This model-checking algorithm takes
  benefit from a parallel state space construction algorithm, which we
  described in a previous work, and shares the same basic set of
  principles: there are no assumptions on the models that can be
  analyzed; no restrictions on the way states are distributed; and no
  restrictions on the way work is shared among processors. We evaluate
  the performance of different versions of our algorithm and compare
  our results with those obtained using other parallel model checking
  tools. One of the most novel contributions of this work is to study a
  space-efficient variant for CTL model-checking that does not require
  to store the whole transition graph but that operates, instead, on a
  reverse spanning tree.
\end{abstract}

\section{Introduction}
\label{sec:introduction_mc}

Several model-checking methods address the state-explosion problem
from a purely algorithmic perspective, for instance with the use of
abstractions on the set of states (such as stubborn sets or
symmetries) or symbolic techniques. Despite the fact that considerable
progress has been made at the theoretical level, there are still
classes of systems that cannot benefit from these advanced methods,
like for example models that rely on real time constraints or on
dynamic priorities. In this case, it is interesting to take benefit
from the computation power---and increased amount of primary
memory---provided by multi-processor and multi-core computers in order
to handle very large state spaces.

In this paper, we propose a parallel algorithm for local, on the fly,
model checking of a fragment of CTL that is well-suited for modern,
multi-core architectures. We target shared-memory computers with a
moderate number of cores (say 4 to 64) operating on a large shared
memory space (typically from \unit{16}{GB} to \unit{1}{TB} of RAM). This
description fits many available mid-range servers, but is also quite
close to tomorrow's mainstream desktop computers.

Our model-checking algorithm takes benefit from a parallel state space
construction algorithm defined in a previous
work~\cite{t._saad_mixed_2011} and share the same principles. First,
we make no specific assumptions on the models that can be analyzed (we
only assume that we know how to compute the successors of a given
state and test them for equality). Second, we put no restrictions on
the way states are distributed: in our solution, every process keeps a
local share of the global state space and we do not rely on an
a-priori static partition of the states. Finally, we put no
restrictions on the way work is shared among processors; this means
that our algorithm plays nicely with traditional work-sharing
techniques, such as work-stealing or stack-slicing. 

In this paper, we extend this state space construction algorithm by adding model
checking capabilities. While the leading parallel model-checking tools
are based on LTL model-checking,
(see Sect.~\ref{sec:comparison_mc_divine})
we advocate the use of CTL.
The choice of CTL derives from a number of desirable properties that
we want for our parallel algorithm. 

First, it must take advantage of parallelism and be compatible with
our parallel state space generator. For this reason, the logic should
preferably be {\em branching time} rather than linear time since model
checking algorithms for linear time logics are strongly tied to
depth-first-search (dfs) exploration techniques; and that dfs
algorithms are ``inherently sequential'' (they belong to the class of
P-complete
problems~\cite{reif_depth-first_1985,greenlaw_limits_1995}). Parallel
techniques for LTL model checking have been quite
investigated however~\cite{BBCR10,laarman_multi-core_2011} so we
will compare our approach with these.

Secondly, we want an {\em on-the-fly} algorithm; only states essential
for answering the model-checking problem should be enumerated. For
this reason, model checking should be {\em local} rather than global;
properties will be interpreted at the initial state rather than at
all states.

Next, because the full state space may have to be enumerated
(e.g. when checking a property that is true), we want it to be
space-efficient.  Hence, we shall accept that a small amount of
information is recomputed every time it is needed rather than kept in
storage.

For all these reasons, we decided to select a fragment of Computation
Tree Logic (CTL) with state subformulas restricted to atomic
propositions. This fragment strictly includes the logic used by
popular tools like Uppaal. Though obviously less expressive than CTL,
it implements a good trade-off between expressiveness and
cost of verification when used to check large state spaces. While not
yet implemented in our tool, it is possible to adapt our algorithm
to model-check full CTL; more expressive logics will be considered
in further work.

\vspace{4pt}{\noindent \em Contributions.}
We follow the classical approach of Clarke and
Emerson~\cite{kozen_design_1982} for CTL model-checking.  During
model-checking, we label each state of the system with the subformulas
that are true at this given state. Labels are computed iteratively
until we reach a fix-point, that is until we cannot add new labels.
We consider two variants of this algorithm that differ by the amount
of information on the transition relation that is stored.  Both
variants have two passes: a \emph{forward pass} performs a constrained
exploration of the state graph in which we start labeling each state
with local information; a second, \emph{backward pass}, propagates
information towards the root of the state graph and checks if the
resulting graph admits an infinite path.

In the first version of the algorithm, that we call \emph{RG} (for
reverse graph), we assume that, for every reachable state, we have a
constant time access to the list of all its ``parents''. In other words,
we store the reverse transition relation of the state space. Algorithm
RG is simply a parallel version of the algorithm
in~\cite{kozen_design_1982} that uses our parallel state space
construction method. Our experimental results show that, even with
this simple approach, we obtain a very good parallel implementation
(that is with a good speedup) and a very good model-checking tool
(that is with a good execution time when compared with other tools on
a similar setup).

In the second version, we assume that we have direct access to only
one of the parents, meaning that we may have to recompute some
transitions dynamically.  We call this second version \emph{RPG}, for
\emph{reverse parental graph}. The advantage of the RPG version is to
save memory space. Indeed, if we use the symbol $S$ to denote the
number of reachable states, the RG algorithm as a space complexity in
the order of $O(S^2)$ in the worst-case, while it is of the order of
$O(S)$ for the RPG version. We show, in our benchmarks, that the RPG
version allows to compute bigger examples without sacrificing
execution time.

\vspace{4pt}{\noindent \em Outline of the paper.}
In the next section, we summarize the parallel state space generation
algorithm~\cite{t._saad_mixed_2011} that is used in our work. The
model checking algorithms are defined in Sect.~\ref{sec:alg_mclcd}
using pseudo-code, while our parallel implementation is described in
Sect.~\ref{sec:experiments_mc}. Before concluding, we discuss the
related work and compare our performances with the DiVinE\cite{BBCR10}
tool, a state of the art parallel model checker for LTL.

\section{Parallel State Space Generation}
\label{sec:stategen}

State space generation is often a preliminary step for model checking
behavioral formulas. This is a very basic operation: take a state
that has not been explored;
compute its successors and
check if they have already been found before; iterate until there is
no more new state to explore. Hence, a key point for performance is to
use an efficient data structure for storing the set of generated
states and for testing membership in this
set. In~\cite{t._saad_mixed_2011}, we propose an algorithm for
parallel state space construction based on an original concurrent data
structure, called a \emph{localization table} (\LT), that aims at
improving spatial and temporal balance.

This approach is close in spirit to algorithms based on distributed
hash tables, with the distinction that states are dynamically assigned
to processors; i.e. we do not rely on an a-priori static partition of
the state space. In our solution, every process keeps a share of the
global state space in a local data structure. Data distribution and
coordination between processes is made through the \LT, that is a
lockless, thread-safe data structure. The localization table is used
to dynamically assign newly discovered states and can be queried to
return the identity of the processor that owns a given state. With
this approach, we are able to consolidate a network of local state
repositories into an (abstract) distributed one without sacrificing
memory affinity---data that are ``logically connected'' and physically
close to each other---and without incurring performance costs
associated to the use of locks to ensure data consistency.

The performance of our state space construction algorithm was
evaluated on different benchmarks and compared with the results
obtained using other solutions proposed in the literature. A first
implementation of our algorithm showed promising results as we
observed performances that are consistently better---both in terms of
absolute speedup and memory footprint---than with other parallel
algorithms. For example, this algorithm does consistently better than
algorithms based on the use of static partitioning or than a similar
approach based on the concurrent hash map implementation provided in
the Intel Threading Building Blocks (TBB) library.

State space generation has a direct impact on the performance of the
model-checking algorithm. For one thing, state space generation alone
is enough to model-check reachability properties (of the form $A \Box
(\phi)$). Moreover, for more complicated properties (see our benchmark results
in Sect.~\ref{sec:experiments_mc}), the time needed
to explore the state space still makes up a big part of the model
checking time.

\section{Parallel Model Checking for a CTL Fragment}
\label{sec:alg_mclcd}

We build our model checking algorithm on top of the parallel state
space generation algorithm of~\cite{t._saad_mixed_2011}, described in
the previous section. Our other design choices follow from our goal
to target models with very large state spaces. More particularly, we
choose to restrict ourselves to a fragment of CTL and to disallow the nesting of
operators; that is, every subformula---denoted $\phi, \psi,
\dots$---is a (boolean composition of) atomic propositions.

The logic used for model-checking essentially relies on three
operators: \emph{Exist Until} (EU), $E\,(\psi \cup \phi)$, that is
true if there exists a trace (a path) in the state space such that
$\psi$ has to hold until, at some position, $\phi$ holds; Always Until
(AU), $A\, (\psi \cup \phi)$, that is true if the ``until condition''
holds on every trace; and finally the \emph{leadsto} formula, $\psi
\leadsto \phi$, that is true if, for every trace, whenever $\psi$
holds then necessarily $\phi$ will hold later. The last property can
be expressed as $A \Box(\neg\, \psi \vee A\Diamond (\phi))$ in CTL.
From the interpretation given in Table~\ref{table:table_of_formulas},
we see that these operators define an expressive
fragment of CTL (and also LTL).

Model-checking procedures for these operators will be described in
Sections~\ref{sec:check-eu-prop} to \ref{sec:check-leadsto-prop}.
In our implementation, we consider two variants---RG and RPG---of the
algorithms.  Both versions are based on two elementary phases: (1) a
forward constrained exploration of the state graph using the state
space construction discussed in Sect~\ref{sec:stategen}; followed by
(2) a backward traversal and label propagation phase ensuring that the
resulting graph is acyclic. 

The backward traversal phase is only needed for {AU} and {leadsto}
formulas, since checking EU formulas amounts to performing a
constrained exploration of the state space (for instance, the formula
$A\Box(\phi)$ is true if no state satisfies $\neg\phi$, which can be
checked during the exploration phase).  Consequently, our algorithm is
 not completely on-the-fly for these cases because the presence of a cycle 
is detected after the (constrained) state space is constructed, delaying the 
discovery of an invalid path. The last column of Table~\ref{table:table_of_formulas} indicates, for each formula,
whether the backward phase is necessary.

\begin{figure}[]
\vspace*{-.3cm}
 	\centering
	\begin{tabular}{|c|c|c|c|}
          \hline 
          Formulas 	& Interpretation in CTL		& Forward & Backward 		\\ \hline 
          $E~ (\psi \cup \phi)$ & $E~ (\psi \cup \phi)$	&	x & 			\\ \hline 
          $A~ (\psi \cup \phi)$ & $A~ (\psi \cup \phi)$	&	x & x			\\ \hline 
          $E \Diamond (\phi)$ 	& $E~ (\mathrm{True} \cup \phi)$	& x & 		\\ \hline 
          $A \Diamond (\phi)$ 	& $A~ (\mathrm{True} \cup \phi)$	& x & x		\\ \hline 
          $E \Box (\phi)$ 	& $\neg A \Diamond  (\neg \phi)$	& x & x		\\ \hline 
          $A \Box (\phi)$ 	& $\neg E \Diamond  (\neg \phi)$	& x & 		\\ \hline  
          $\psi \rightsquigarrow \phi$ & $A\Box( \neg \psi \vee A\Diamond \phi)$ & x &x 	\\ \hline
          $A \Box A \Diamond (\phi)$ & $ {true} \rightsquigarrow \phi$ &	x &x 	        \\ \hline
	\end{tabular}
	\caption{List of Supported Formulas.\label{table:table_of_formulas}}
\vspace*{-.5cm}
\end{figure}

\subsection{Concepts and Notations} 

We assume that we perform model-checking on a Kripke System
$\KS(S,R,s_0)$. We will use, interchangeably, the notation \KS for the
Kripke structure $(S,R,s_0)$ and $G$ for the directed graph $G(S,R)$,
also called the \emph{state graph}. In the RPG version of our
algorithm, we make use of the \emph{Parental Graph} of a Kripke
System, that is a reverse spanning tree of the (currently computed)
state graph.

\begin{definition}[Parental Graph]
\label{defn:pg}
The directed graph, $PG(V_p,E_p)$, is a parental graph of $G(V,E)$ if:
(1) $PG$ if a subgraph of $G$ that has the same vertices, that is $V_p
= V$ and $E_p \subseteq E$, and (2) for every vertex $v \in V$, if $v$
is not the root of $G$ then $v$ has an in-degree of one in $PG$.
\end{definition}

A simple way to obtain a parental graph, $PG$, when exploring the
state graph, $G$, is to keep for every state, $s$, a vertex to the
state in $G$ that was used to generate $s$ (and forget the others).
The parental graph has nice properties. If $PG$ is a parental graph of
$G$ and $G$ is acyclic then so is $PG$.  Moreover, the set of leaves
of $PG$ subsumes that of $G$; a leaf of $G$ is necessarily a leaf of
$PG$.

In the remainder of the text, the expression $|S|$ is used to denote
the cardinality of $S$ (the number of reachable states), while $|R|$
is the number of transitions.
We assume that every state $s \in S$ is labeled with a value, denoted
$\suc(s)$, that records the out-degree of $s$ in \KS. The value of
$\suc(s)$ is set during the forward exploration phase. Initially,
$\suc(s)$ is the cardinality of the set of successors of $s$ in \KS,
that is $\suc(s) = \left|\{ s' \,|\, s \mathrel{R} s' \}\right|$. We
decrement this label during the backward traversal of the state graph;
when the value of $\suc(s)$ reaches zero, we say that $s$ is
\emph{cleared} from the state graph. In our pseudo-code, we use the
expression $\suc(s).\dec()$ to decrement the value of the label $\suc$
for the state $s$ in \KS, and the expression $\suc(s).\set(i)$ to set
the label of $s$ to some integer value $i$.

When we deal with the reverse parental graph version of our algorithm,
we assume that we implicitly work with one particular parental graph
of \KS, denoted \PKS. In this case, we assume that every state $s \in
S$ is also labeled with a value, denoted $\sons(s)$, that records the
out-degree of $s$ in \KS. We also label each state $s \in S$ with a
state, denoted $\father(s)$, that is the (unique) predecessor of $s$
in \PKS. (The label $\father(s)$ makes sense only if $s$ is not the
initial state, $s_0$, of \KS.) Initially, the value of $\sons(s)$ is
set to zero. The label will be incremented during the forward
exploration, when we build \PKS (that is, we select the transitions
from \KS that will be stored in \PKS). This operation is denoted
$\sons(s).\inc()$ in our pseudo-code. We will decrement the value of
$\sons(s)$ during the backward traversal phase.

\subsection{Checking EU properties}
\label{sec:check-eu-prop}

Checking \emph{EU} properties for the initial state is standard,
except that we perform the forward phase concurrently, on all
states. this can be done on the fly in a single, forward pass.  To
check the formula $E\, (\psi \cup \phi)$, we explore the state space
until a state is found such that either (1) $\phi$ holds and no more
state has to be explored, or (2) $\neg\psi \wedge \neg\phi$ holds.  In
the first case, the algorithm reports success; all states obeying
$\phi$ terminate a (possibly empty) path of states rooted at the
initial state and all obeying $\psi$.  In the second case, we have
found a counter-example; a state obeying neither $\psi$ nor $\phi$,
meaning that the property is false at the initial state. The check
function is the same for the two versions of our algorithm, whether
based on the reverse graph or the reverse parental graph data
structure.

\begin{lstlisting}[mathescape=true,%
 caption=Algorithm for the formula $A\, (\psi \cup \phi)$---function check\_a,%
 label=alg:pseudocode-check-a, language=pseudo]
function BOOL check_a($\psi$ : pred, $\phi$ : pred, $s_0$ : state)
   Stack A = new Stack($\emptyset$) ;
   // Start with the forward exploration
   if forward_check_a($\psi$ , $\phi$, $s_0$, A) then // If all forward constraints are respected
      return backward_check_a($s_0$, A)       //start the backward phase
   else return FALSE // We found a problem during the forward exploration
\end{lstlisting}

\vspace*{-.3cm}

\subsection{Checking {AU} Properties}
\label{sec:model-check-liven}

For checking the formula $A\, (\psi \cup \phi)$, as for \emph{EU}
properties, we stop exploring a path when we find a state such that
(1) $\phi$ holds or (2) $\neg\psi \wedge \neg\phi$ holds. If an
occurrence of case (2) is found, we have a counter-example to the
property false. Otherwise, we start backward traversal phase in order
to detect cycles. Indeed, the property $A\, (\psi \cup \phi)$ is false
if there is an infinite path of states (starting from $s_0$) that all
obey $\psi$. We call this second phase the \emph{clearing phase},
because it consists in recursively removing leaf nodes from the
graph. This process ends either when the only remaining state is the
initial state (meaning that the property is true), or when no states
with out-degree zero can be found (in which case we know that there is
a cycle). The validity of this method follows from the fact that a
finite Directed Acyclic Graph (DAG) has at least one leaf.

We give the pseudo-code for checking $A\, (\psi \cup \phi)$ in
Listing~\ref{alg:pseudocode-check-a}.  The inputs are the atomic
properties $\psi$ and $\phi$ and the initial state $s_0$. The
algorithm makes use of a stack A to collect the states at which $\phi$
holds during the forward exploration phase. The procedure uses two
auxiliary functions, forward\_check\_a and backward\_check\_a, that
depends on the version of the algorithm. We start by defining these
helper functions for the Reverse Graph version.

\vspace{4pt}{\noindent \em Algorithm for the Reverse Graph version (RG).} We give the
pseudo-code for the function {forward\_check\_a} (for the RG version)
in Listing~\ref{alg:pseudocode-backward-check-a-rg}. The last
parameter of this function, A, is a stack that is used to collect the
leaf nodes of the state graph, that is the states where $\phi$ holds.
These states are the starting points in our backward traversal of the
graph.

During the forward exploration phase (function forward\_check\_a)
each state $s$ is labelled with its number of successors in the initial
state graph (the Kripke structure). During the backward traversal
phase (function backward\_check\_a), 
this label is decremented each time a successor of $s$ is removed;
decrementations are done in parallel. Intuitively, a
state can be removed as soon as it is cleared. We never actually
remove a state from the graph. Instead, when a processor changes the
label of a state $s$ to $0$,
we also decrement the labels of
all the parents of $s$ in the graph. Hence the choice of storing the
reverse of the transition function in the data structure.

In the function backward\_check\_a, see
Listing~\ref{alg:pseudocode-backward-check-a-rg}, we start by clearing
all the states in A which are, by construction, the states $s$ such that
$\suc(s)$ is nil. When a state is cleared, we decrement the labels of
all its parents ($suc(s').dec()$) and check which ones can be cleared
($suc(s')==0$). The algorithm stops if the initial state, $s_0$, can
be cleared or if there are no more states to update.

\begin{lstlisting}[%
 caption=Forward and backward exploration for $A\, (\psi \cup \phi)$ with Reverse Graph,%
 label=alg:pseudocode-backward-check-a-rg, language=pseudo, mathescape=true]
function BOOL forward_check_a($\psi$ : pred, $\phi$ : pred, $s_0$ : state, A : Stack)
   Set S = new Set($s_0$) ; Stack W = new Stack($s_0$) ;
   while (W is not empty) do
      s = W.pop();
      if (s $\vDash \phi$) then
         suc(s).set(0) ; // We clear state s from KS
         A.push(s)
      elsif (s $\vDash \psi$) then // We tag s with its number of successors
         suc(s).set(number of successors of s in KS) ;
         if (suc(s) == 0) // Check if s is not a dead state
            return FALSE
         forall s' successor of s in KS do // and continue the exploration
            if (s' NotIn S) then
               S = S $\cup$ {s'} ; // s' is a new state 
               W.push(s')
      else return FALSE
   return TRUE

function BOOL backward_check_a($s_0$ : state, A : Stack)
   while (A is not empty) do
      s = A.pop() ;
      if (s == $s_0$) then // The property is true if 
         return TRUE    // we reach the initial state
      forall s' parent of s in KS do	// Otherwise we check if the
         suc(s').dec() ;               // predecessors of s can be cleared
         if (suc(s') == 0) then A.push(s')
   return FALSE
\end{lstlisting}

\vspace*{-.3cm}

\vspace{4pt}{\noindent \em Algorithm for the Reverse Parental Graph version (RPG).}
The function for the RPG version is only slightly more complicated,
because we need to recompute some successors in the transition
relation: we can only access one of the parents of a state in constant
time (which we call the {\em father} of the state). The pseudo-code
for the forward exploration phase (function {forward\_check\_a}) is
essentially the same as in
Listing~\ref{alg:pseudocode-backward-check-a-rg}; this is why it is
omitted here. Compared to the RG version, we only need to add two
additional statements when adding a new state (around line 15 in
Listing~\ref{alg:pseudocode-backward-check-a-rg}): assuming that the state
$s$ is generated from a state $s'$, we set the value of the father for
the newly generated state ($father(s).set(s')$) and increment the
number of sons of the father $sons(s).inc()$). This information is
used during the backward traversal to track non cleared leaves.

We give the pseudo-code for the backward traversal phase in
Listing~\ref{alg:pseudocode-backward-check-a-rpg}. During this phase,
we follow the parental graph structure to ``propagate'' the cleared
states toward the root of the state graph. The algorithm alternates
between two behaviors, \textit{clearing} and \textit{collecting}. The
\textit{clearing} behavior is similar to the pseudo-code for the RG
algorithm, with the difference that we decrement only the father of a
state and not all the predecessors. When there are no more labels to
decrement---and if the root state is not yet cleared---the algorithm
starts looking for states that can be cleared. For this, we test all
the states $s$ such that $sons(s)==0$; that is, such that all the sons
of $s$ have been cleared (in the parental graph). In this case, to
check if $s$ can be cleared, we have to recompute all its successors
in \KS and check whether they have also been cleared (if their $\suc$
label is zero).

\begin{lstlisting}[%
    caption=Backward exploration for $A\, (\psi \cup \phi)$ with Reverse Parental Graph,%
    label=alg:pseudocode-backward-check-a-rpg, language=pseudo, mathescape=true]
function BOOL backward_check_a($s_0$ : state, A : Stack)
   over = FALSE
   while (not over)
      while (A is not empty) do
         //Clearing 
         s = A.pop() ;
         if (s == $s_0$) then  // The property is true if 
            return TRUE    // we reach the initial state
         s' = father(s) ;	// Otherwise we check if 
         sons(s').dec() ;  // the father of s can be cleared
         suc(s').dec() ;
         if (suc(s') == 0) then A.push(s')
      //Collecting: if we have no more states to clear in A we try to find 
      // candidates among the states with no children in \PKS
      forall s such that sons(s) $=$ 0 and suc(s) $\neq$ 0 in KS do
         if test(s) then
            suc(s).set(0) ;
            A.push(s) 
      if (A is empty) then
         over = TRUE //No good candidate was found, end backward search
   return FALSE

function BOOL test(s : state)
   forall s' successor of s in KS do
      if suc(s') != 0 then
         return FALSE // at least one successor is not cleared
   return TRUE
\end{lstlisting}

The advantage of this strategy is that we do not have to consider all
the states in the graph but just a subset of them.  Indeed, we know that
if \KS is a acyclic (is a DAG) then $PG$ has at least one leaf that is
also a leaf in $G$~\cite{thesis_SAAD_2011}. Hence, this subset is
enough to test the presence of a cycle. Conversely, the drawback of this approach is that we may try to clear the same vertex several times, which may be time consuming.

\subsection{Checking Leadsto Properties}
\label{sec:check-leadsto-prop}

To check the formula $\psi \rightsquigarrow \phi$, we need to prove
that no cycle can be reached from a state where $\psi$ holds, without
first reaching a state where $\phi$ holds. Indeed, otherwise, we can find an infinite path   where

\begin{wrapfigure}[16]{R}{0.4\textwidth}
\vspace{-20pt}
  \centering
  \includegraphics[width=0.4\textwidth]{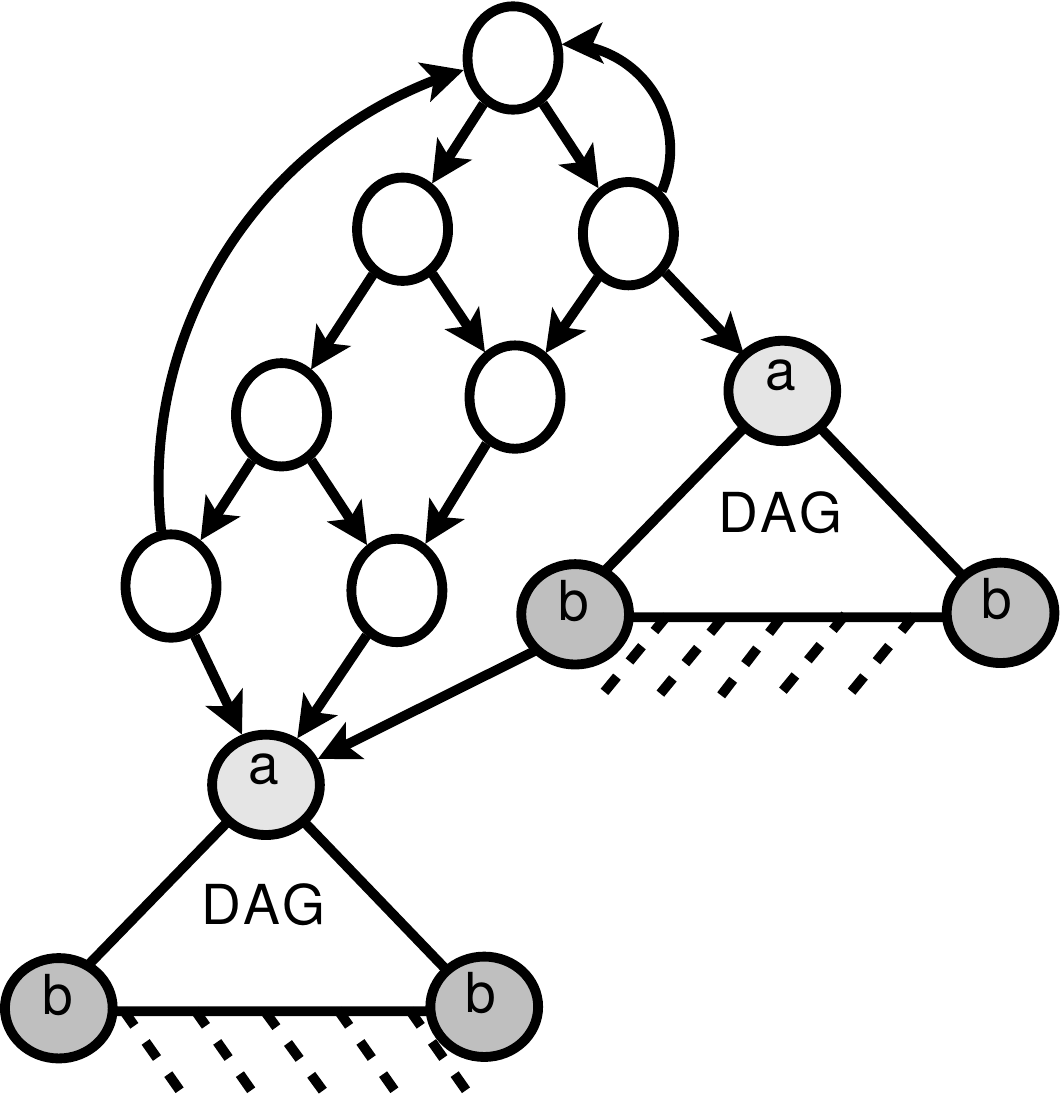}
  \caption{Formula $a \rightsquigarrow b$.}
  \label{fig:dag_leadsto}
  \noindent
\end{wrapfigure}

\noindent  $\phi$ never holds after an
occurrence of $\psi$. Figure~\ref{fig:dag_leadsto} shows an example of
graph for which the formula is valid.

This observation underlines the link between checking the formula
$\psi \rightsquigarrow \phi$ locally---for the initial state---and
checking the validity of $A\Diamond (\phi)$ globally---at every state
where $\psi$ holds. As a consequence, we can use an approach similar
to the one used for {AU} properties in the previous section. The main
difference is that, instead of clearing the initial state, we have to
clear all the states where $\psi$ holds. Hence, the pseudo-code for
the leadsto formulas is similar to that of {AU} formulas (this is why
it is omitted here), the main difference is in the termination
condition: the function returns true if all the states where $\psi$
holds are cleared.

\subsection{Correctness and Complexity of our Algorithms}
\label{sec:theoretical_mc}
\enlargethispage{\baselineskip}

Proofs of correctness (termination, completeness and soundness) and a
precise study of the complexity of our algorithms can be found
in~\cite{thesis_SAAD_2011}. We just discuss here the worst-case
complexity in the sequential case, and for formulas $A\,
(\psi \cup \phi)$. The results for this case can be generalized to our
whole logic. (Inside asymptotic notations, we use the symbols $S$ and
$R$ when we really mean $|S|$ and $|R|$.)

The algorithm given in Sect.~\ref{sec:model-check-liven} may inspect
every state in the Kripke Structure \KS and, for every transition, it
may update one label. Therefore, its worst-case time complexity is in the order of $O(S + R)$ for the RG algorithm. The
complexity is higher in the RPG version since, for each altered state,
we may have to recompute the successors for all the reachable states
$s$ such that $\sons(s)$ is nil. Hence, solving a simple recurrence, we
can prove that the time complexity is in the order of $O(S \cdot (R -
S))$ for the RPG version.  Since the number of transitions in \KS is
bounded by $|S|^2$, we obtain a complexity in the order of $O(S^2)$
for the RG version and of $O(S^3)$ for the RPG variant. Concerning the
space complexity, the RPG version is in the order of $O(S)$, while the
RG version is linear in the size of the graph, that is in the order of
$O(S + R)$ (or $O(S^2)$). 

We show in our experiments that the decision to favor
``space-efficiency'' (in the case of the RPG version) is quite
interesting. In particular, on some examples, the RPG version may run
faster than the RG version because it needs to ``write less
information'' in main memory, an effect that is not visible if we only
look at the theoretical complexity. More importantly, memory is one of
the key resources used during model-checking. Indeed, it is common to exhaust
the available memory during verification.

\section{Implementation and Experimental Results}
\label{sec:experiments_mc}

In the code presented in Sect.~\ref{sec:alg_mclcd}, no 
underlying computational model was make precise. The code can
be easily adapted to a Parallel RAM model, following a Single Program
Multiple Data (SPMD) programming style. In this section, we discuss
the details surrounding the parallel implementation of our algorithms,
then report on a set of experiments performed to evaluate their
effectiveness.

\subsection{Parallel Implementation of our Algorithm}
\enlargethispage{\baselineskip}

In a SPMD context, all processing units will execute the same
functions (the one defined in
Listings~\ref{alg:pseudocode-check-a}--\ref{alg:pseudocode-backward-check-a-rpg}).
Following this approach, the (forward) exploration phase and the
(backward) cycle detection phase can both be easily
parallelized. Then, for the model-checking function themselves---for
instance the function check\_a---we only need to synchronize the
termination of the forward exploration with the start of the backward
label propagation. At each point, a processing unit can terminate the
model-checking process if it can prove (or disprove) the validity of
the formula before the end of the exploration phase. Actually, most of
the burden of parallelizing our algorithm is hidden inside the use of
our specialized, lock-free data structures.

We consider a shared memory architecture where all processing units
share the state space (using the mixed approach presented
in~\cite{t._saad_mixed_2011}) and where the working stacks are
partially distributed (such as the stacks W and A used in our
pseudo-code). For most of our pseudo-code, it is enough to rely on
atomic “compare and swap” primitives to protect from parallel data
races and other synchronization issues; typically, compare-and-swap
primitives will be used when we need to test the value of a label or
when we need to update the label of a state (for instance with
expressions like $\sons(s).\dec()$). Together with the
compare-and-swap primitive, we use our combination of distributed,
local hash tables with a concurrent localization table to store and
manage the state space.

For the RG version of the algorithm, we can ensure the consistency of
our algorithm by protecting all the operations that manipulate a state
label. (We made sure, in our pseudo-code, that every operation only
affects one state at a time.) The parallel version of RPG is a bit
more involved. This problem is related to the behavior of the
\emph{collecting} operations of the backward exploration (see the
comment on line 14 of
Listing~\ref{alg:pseudocode-backward-check-a-rpg})---and in particular
the function test---that needs to check all the successors of a state
to see if they are cleared. First, this code is not atomic and it is
not practical to put it inside a critical section (it would require a
mutex for every state). If two processors collect the same state,
then the father of this state could be decremented twice, during the
\emph{clearing} operations. Second, \textit{collecting} must be
performed after all processors have finished the \textit{clearing}
operations, otherwise the algorithm may end prematurely
(see~\cite{thesis_SAAD_2011} for a complete explanation.)  We solve
the concurrency issues for the RPG version through the synchronization
of all processors before both \textit{clearing} and
\textit{collecting}.  Then, we take advantage of our distributed local
state repositories to avoid problems due to concurrent access; each
processor can perform the \textit{collecting} operations only over the
states that it owns. Finally, we use a work-stealing strategy
(see~\cite{t._saad_mixed_2011}) to balance the work-load between the
different phases of our algorithm; for instance, whenever a thread has
no more state to clear, it tries to ``steal'' non-cleared states from
other processors.

\subsection{Experimental Results}

We have implemented the parallel versions of our model-checking
algorithm and evaluated their performances on several benchmarks.
Experimental results presented in this section were obtained on a Sun
Fire x4600 M2 Server configured with 8 dual core opteron processors
and \unit[208]{GB} of RAM memory, running the Solaris 10 operating
system. (The complete set of experiments can be found 
in~\cite{thesis_SAAD_2011}.)  We give results obtained on 8 classical
models---a Token Ring protocol; the Peg-Solitaire board-game;
\dots---with a mix of valid and invalid properties. We experimented
with all the formulas: reachability ($E\Diamond \phi$), safety ($A\Box
\phi$ and $E\Box \phi$), liveness ($A\Diamond \phi$) and leadsto
($\psi \rightsquigarrow \phi$).

\vspace{4pt}{\noindent \em Speedup Analysis:} we study the relative speedup and the
execution time for our algorithms. In addition, we also give the
separate speedup obtained in each phase of the algorithm---during the
exploration (forward) and cycle detection (backward) phases---in order
to better analyze the advantages of our approach.

\begin{figure}[!htb]
\vspace*{-.6cm}
  \centering	
  \begin{tabular}{c}
   \begin{minipage}{\linewidth}
      \begin{tabular}{cc}
      \begin{minipage}{0.65\linewidth}
	      \centering
	      \includegraphics[width=\linewidth]{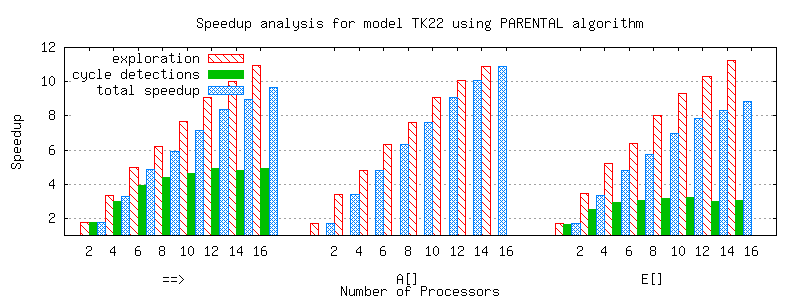}
      \end{minipage}
	    &
      \begin{minipage}{0.35\linewidth}
	      \centering
	      \includegraphics[width=\linewidth]{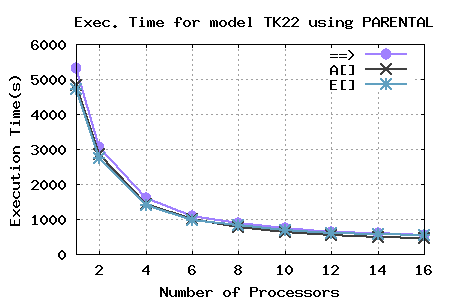}
      \end{minipage}
					  \\
					  \begin{minipage}{0.65\linewidth}
	      \centering
	      \includegraphics[width=\linewidth]{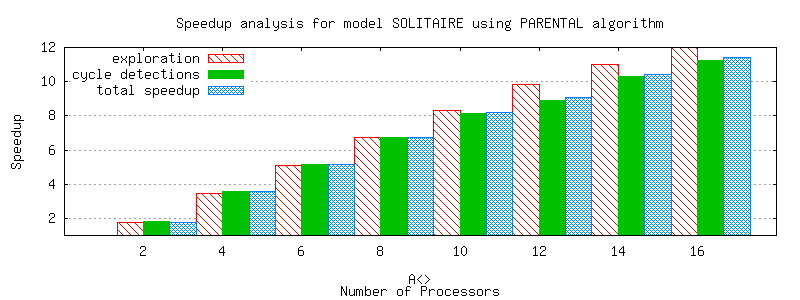}
      \end{minipage}
	    &
      \begin{minipage}{0.35\linewidth}
	      \centering
	      \includegraphics[width=\linewidth]{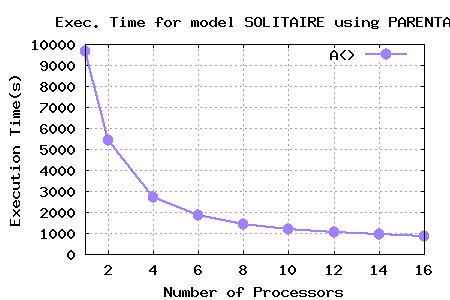}
      \end{minipage}
					  \\
					  a) Speedup & b) Execution Time
    \end{tabular} 
    \caption{Speedup and execution time analysis for Token Ring and Solitaire models.\label{fig:speedup_execution_time}}
   \end{minipage}
    \\
    \begin{minipage}{\linewidth}
      \centering
      \begin{tabular}{cc}    
	      \begin{minipage}{0.5\linewidth}
		\centering
		\includegraphics[width=.95\linewidth]{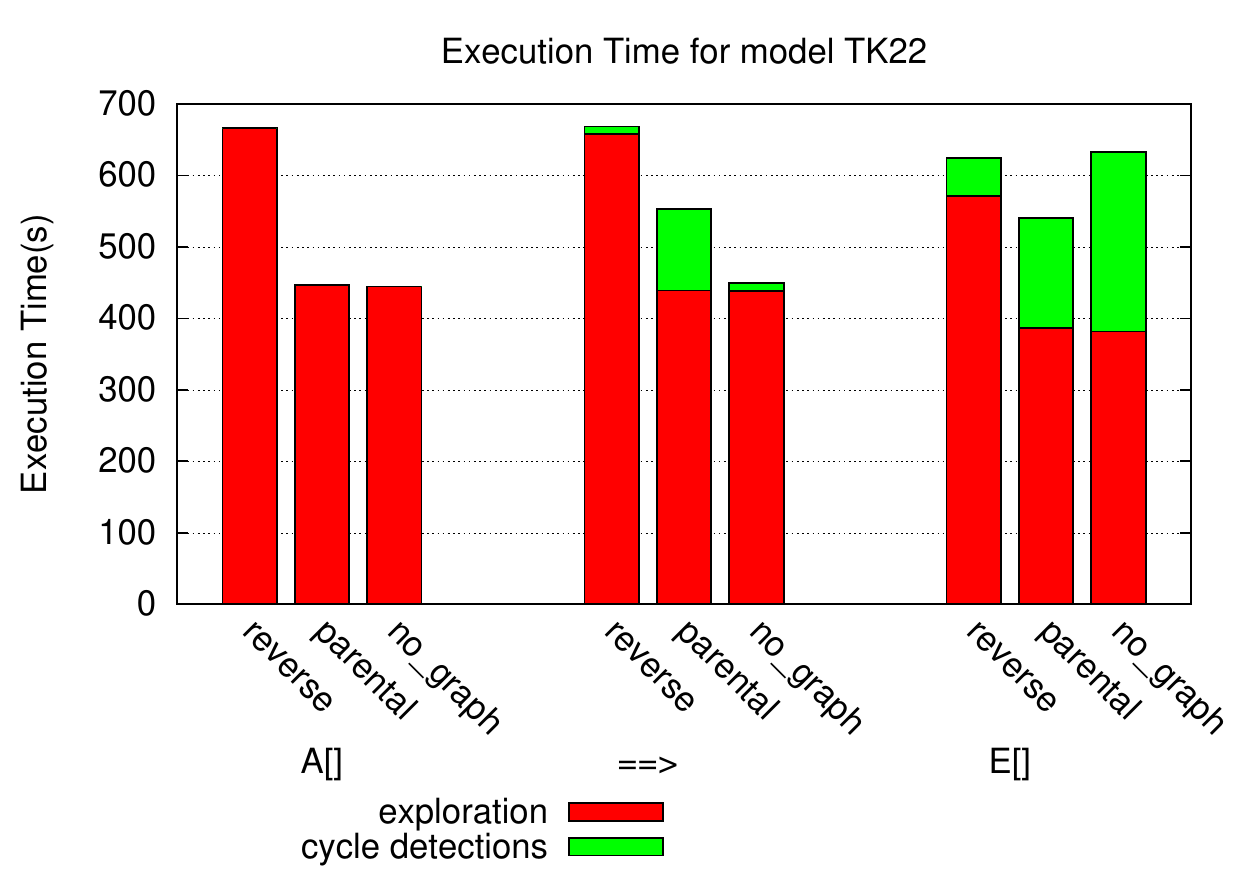}
	      \end{minipage}
	      &
	      \begin{minipage}{0.5\linewidth}
		\centering
		\includegraphics[width=.95\linewidth]{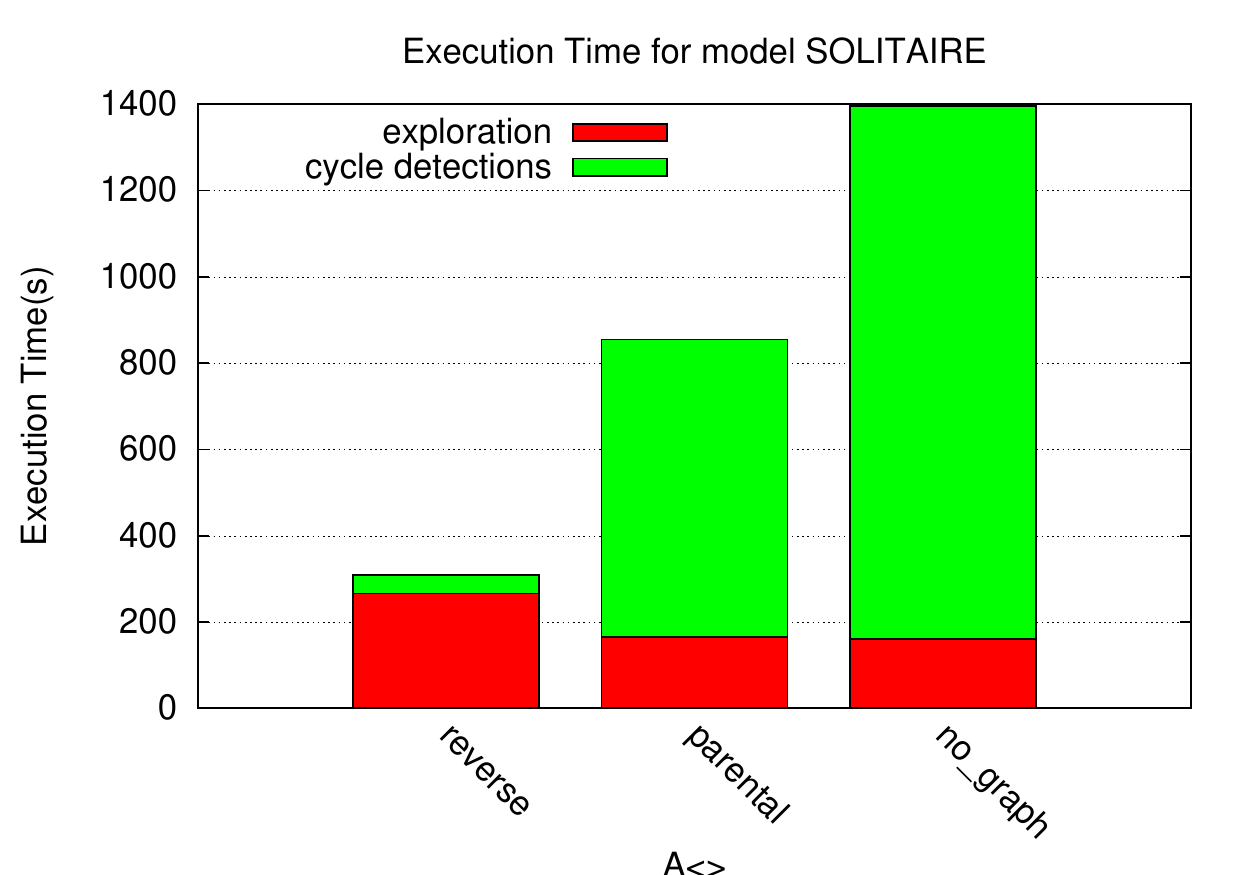}
	      \end{minipage}
      \end{tabular} 
	\caption{Comparison with a Standard Algorithm.\label{fig:bar_chart_analysis}}
   \end{minipage}
  \end{tabular}  
\vspace*{-.6cm}
\end{figure}

Figure~\ref{fig:speedup_execution_time} shows speedup analysis for
the RPG version of our algorithm. We only show the results for two
models---a Token Ring with 22 bases (TK22) and a Solitaire game with
33 pegs---since they are representative of the results obtained with
our complete benchmark. These models have different execution profiles
which impacts significantly the overall performance. The main
difference is the time spent in the backward traversal phase.
Figure~\ref{fig:bar_chart_analysis} shows a series of bar charts putting
in evidence the time required for each phase of the algorithm
(exploration and cycle detection). In addition, we compare our
approach (RG and RPG) with a third algorithm (NO\_GRAPH) that uses the
same code as RG but recomputes predecessor states instead of storing
them (this is possible only because, for this particular benchmark, we
know how to compute the predecessors of a state). We have observed two
main categories of behaviors in this analysis.

\vspace{2pt}
{\noindent \bf negligible backward traversal:} the time spent in the backward
  exploration phase is negligible compared to the overall execution
  time (e.g. model TK22 in Fig.~\ref{fig:speedup_execution_time},
  \ref{fig:bar_chart_analysis}). This is the case, for instance,
  if the property is false and the cycle detection phase terminates
  early. In this category of experiments, there are no significant
  differences between RG and RPG, mainly because the gain in
  performance during the forward exploration phase outweighs the extra
  work performed during the cycle detection phase;
 
\vspace{2pt}
{\noindent \bf complete backward traversal:} the cycle detection phase needs to
  run through all the state space (e.g. model SOLITAIRE in
  Fig.~\ref{fig:speedup_execution_time}, \ref{fig:bar_chart_analysis}). We observe a significant
  difference in performance between the RG and RPG versions in this
  case. The extra work performed by the RPG version becomes the
  dominant factor.

These experiments confirm that RPG is a good choice when we are
limited by the memory space: although it may require more
computations (in our examples, we may loose a factor of $5$ in
execution time), it can be applied on models that are not tractable
with the RG version because of the space needed to store the
transitions. For instance, for the Peg Solitaire model with 37 pegs
(that has $3.10^9$ states and $3.10^{10}$ transitions), the RPG only
needs \unit[15]{GB} of memory while, with the RG version, we would
need \unit[240]{GB} of memory just to store the transitions.

\section{Related work and Comparisons With Other Tools}
\label{sec:comparison_mc_divine}

Several works address the problem of designing efficient, parallel
model-checking algorithms. Most of the proposals follow an
``automata-theoretic approach'' for LTL model checking. In this
context, the difficulty is to adapt the cycle detection algorithms
(Tarjan or Nested-DFS), which are inherently sequential. Two works
stand out: one with a mature implementation, DiVinE~\cite{BBCR10},
with the \textbf{owcty + map} algorithm; another with a prototype,
named LTSmin, with the \textbf{mc-ndfs}
algorithm~\cite{laarman_multi-core_2011}. They mostly differ by the
algorithm used to detect cycles. 

DiVinE combines two algorithms, \textbf{owcty} and \textbf{map}, that
result in ``a parallel on-the-fly linear algorithm for model
checking weak LTL properties'' (weak LTL properties are those
expressible by an automata that has no cycle with both accepting and
no-accepting states on its path). If the LTL property does not meet
this requirement, the algorithm complexity may be quadratic. The
\textit{multi-core nested DFS} (\textbf{mc-ndfs})
algorithm~\cite{laarman_multi-core_2011} is a recent extension of the
\textbf{swarm}~\cite{holzmann_swarm_2008} distributed algorithm to a
multi-core setting. The authors in~\cite{laarman_multi-core_2011}
propose a multi-core version with the distinction that the storage
state space is shared among all workers in conjunction with some
synchronization mechanisms for the nested search. Even if, in the
worst-case, all the processors may duplicate the same work, this
approach has a linear complexity (given a fixed number of processors).

In contrast with the number of solutions proposed for parallel LTL
model checking, just two specifically target CTL model checking on
shared memory machines: Inggs and Barringer
work~\cite{inggs_ctl*_2006} supports CTL${}^{*}$, while van de Pol and
Weber work~\cite{pol_multi-core_2008} supports the $\mu$-calculus.

\begin{figure}[!htb]
\vspace*{-.4cm}
\centering
\begin{tabular}{c}
  \begin{minipage}{\linewidth}
  \centering
  \begin{tabular}{|c|p{0.55\linewidth}|c|}
    \hline
    Model & Formula & Results\\ \hline
    \multirow{2}{*}{
			\begin{minipage}{.22\linewidth} 
				Anderson (AN)\\ 
        $18\cdotp10^6$ states  
			\end{minipage}} & F1:\verb+(-cs_0) ==>  (cs_0) + 
    & \textit{false}\\   \cline{2-3}
    & F2:\verb+A[]<>(cs_0 or ... or cs_n)+ 
    & \textit{true}\\ \hline
    \multirow{3}{*}{
			\begin{minipage}{.22\linewidth}
				Lamport (LA) \\ 
				$38\cdotp10^6$ states. 
			\end{minipage}} & F1:\verb+(wait_0 and (- cs_0)) ==> (cs_0)+
    & \textit{false}\\ \cline{2-3}
    & F2:\verb+(- cs_0) ==> (cs_0)+
    & \textit{false}\\ \cline{2-3}
    & F3:\verb+A[]<>(cs_0 or ... or cs_n)+ 
    & \textit{true}\\ \hline
    \multirow{2}{*}{
			\begin{minipage}{.22\linewidth}
				Rether (RE) \\
        $4\cdotp10^6$ states 
			\end{minipage}} & 
    F1:\verb+A[]<>(nrt_0)+ 
    & \textit{true}\\  \cline{2-3}
    &  F2:\verb+A[]<>(rt_0)+ 
   & \textit{false}\\ \hline
    \multirow{3}{*}{
			\begin{minipage}{.22\linewidth} 
				Szymanski (SZY) \\
        $2\cdotp10^6$ states .
			\end{minipage}} 
		&F1:\verb+(wait_0 and (- cs_0)) ==> (cs_0)+
    & \textit{false}\\  \cline{2-3}
    & F2:\verb+(- cs_0) ==> (cs_0)+
    & \textit{false}\\  \cline{2-3}
    & F3:\verb+A[]<>(cs_0 or ... or cs_0)+ 
    & \textit{true}\\ \hline
  \end{tabular}
  \caption{Formulas and Models for our Comparison.\label{fig:formulas_benchmark_comp_divine}}
  \end{minipage}
  \\
  \\
  \begin{minipage}{\linewidth}
  \centering
  \begin{tabular}{|c|l|c|c|c|c|c|c|c|c|c|c|}
    \hline
   \multirow{2}{*}{M} & \multirow{2}{*}{~Formula} & \multicolumn{2}{c}{owcty} & \multicolumn{2}{|c}{map} & \multicolumn{2}{|c}{reverse} & \multicolumn{2}{|c|}{parental}\\ \cline{3-10}
 		 & & \begin{minipage}{.08\linewidth}\centering T.(s)\end{minipage}
		 &\begin{minipage}{.085\linewidth}\centering  M.(Gb) \end{minipage}
		 &\begin{minipage}{.08\linewidth}\centering  T.(s) \end{minipage}
		 &\begin{minipage}{.085\linewidth}\centering  M.(Gb) \end{minipage}
		 &\begin{minipage}{.08\linewidth}\centering  T.(s) \end{minipage}
		 &\begin{minipage}{.085\linewidth}\centering  M.(Gb) \end{minipage}
		 &\begin{minipage}{.08\linewidth}\centering  T.(s) \end{minipage}
		 &\begin{minipage}{.085\linewidth}\centering  M.(Gb) \end{minipage}\\ \hline

   \multirow{2}{*}{
			AN
		} 
		& F1:~\textit{false} 
    & 61.3 & 3.3 & 110.2 & 5.5 & 28.8 & 2.8 & 94.4 & 1.8 \\   \cline{2-10}
    & F2:~\textit{true}
    & 79.5 & 7.4 & 110.5 & 4.8 & 26.4 & 2.9 & 50.4 & 1.8 \\ \hline
    \multirow{3}{*}{
			LA
		} 
		& F1:~\textit{false} 
    & 1.6 & 1.1 & 1.4 & 1.1 & 42.4 & 5.1 & 74.2 & 3.3 \\ \cline{2-10}
    & F2:~\textit{false} 
   & 1.4 & 1.1 & 1.7 & 1.2 & 47.6 & 5.6 & 327.2 & 3.6\\ \cline{2-10}
    & F3:~\textit{true}
    & 153.6 & 14.1 & 282.8 & 12.1 & 51.0 & 5.6 & 370.4 & 3.7 \\ \hline
    \multirow{2}{*}{
			RE
		} 
		& F1:~\textit{true}
    & 12.0 & 1.8 & 20.1 & 1.3 & 5.0 & 0.7 & 12.0 & 0.6\\  \cline{2-10}
    & F2:~\textit{false}
    & 13.2 & 1.8 & 1.2 & 0.3 & 3.4 & 0.7 & 7.8 & 0.6 \\ \hline
    \multirow{3}{*}{
			SZY
		} 
		& F1:~\textit{false} 
    & 8.5 & 0.9 & 7.0 & 0.5 & 2.2 & 0.3 & 1.4 & 0.2 \\  \cline{2-10}
    & F2:~\textit{false}
    & 9.8 & 0.9 & 6.6 & .5 & 4.2 & 0.3 & 39.6 & 0.3 \\  \cline{2-10}
    & F3:~\textit{true} 
    & 9.0 & 0.9 & 24.7 & 0.6 & 3.8 & 0.3 & 32.8 & 0.3\\ \hline
 
  \end{tabular}
  \caption{Table of results.\label{fig:results_table}}
  \end{minipage}
  \\
  \begin{minipage}{\linewidth}
		\includegraphics[width=.9\linewidth]{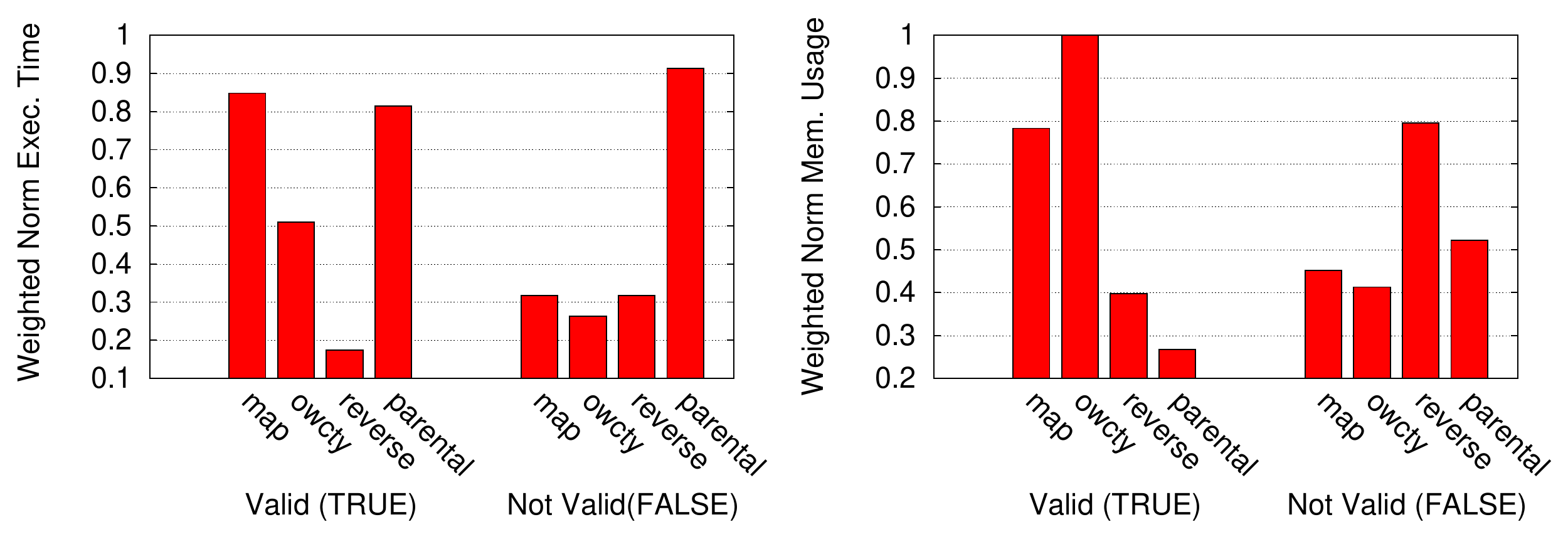}
                \caption{Comparison with divine (reverse = RG, parental = RPG).\label{fig:mc_av_comp_divine_histograms1}}
	\end{minipage}
 
\end{tabular}
\vspace*{-.4cm}
\end{figure}

\vspace{4pt}{\noindent \em Comparison with DiVinE.} We now compare our algorithms
with DiVinE~\cite{BBCR10}, which is the state of the art tool for
parallel model checking of LTL. The results given here have been
obtained with DiVinE 2.5.2, considering only the best results given by
the owcty or map, separately. This benchmark (experimental data and examples 
are available in report \cite{saad_MCLCD_2012})
is based on the set of models
borrowed from DiVinE on which, for a broader comparison, we check both
valid and non valid properties.  Figure
\ref{fig:formulas_benchmark_comp_divine} shows the exact set of models
and formulas that are used.  All experiments were carried out using 16
cores and with an initial hash table sized enough to store all
states. The DiVinE experiments were executed with flag (\textit{-n})
to remove counter-example generation.

Figure \ref{fig:results_table} shows for each model the execution time
(T.) in seconds and the memory peak (M.) (in GB).
Figure~\ref{fig:mc_av_comp_divine_histograms1} summarizes these
results using the normalized weighted sum of the memory footprint and
the execution time, separated for valid and non valid formulas.

Algorithms owcty and map show better overall results when the formula
is not valid (FALSE). By contrast, reverse holds the best execution
time when the formula is valid. Regarding the RPG version of our
algorithm, our results show that it holds the best memory footprint
among all results, it uses on average $2$ to $4$ times less memory
than map and owcty when the formula is valid. In addition, regardless
of its ``cubic'' worst-case complexity, it shows good results when
compared to map and owcty. For instance, it is able to verify a valid
formula on average using $4$ times less memory than owcty with a
limited slow-down ($\approx 1.8$ times slower).

To conclude, for the set of models and formulas used in this
benchmark, both RPG and RG delivered good results when compared to
DiVinE. For instance, RG has a better performance in both time and
memory usage when compared with DiVinE (map and owtcy). Finally, RPG
proved to be the most \textit{space conscious} algorithm---the one to
choose for the biggest models---without sacrificing too much
the execution time.

\section{Conclusion}
\label{sec:conclusion_mc}

We have described ongoing works concerning parallel (enumerative)
model-checking algorithms for finite state systems. We define two
versions of a new model checking algorithm that support an expressive
fragment of both CTL and LTL. These algorithms are based on the
standard, semantic model-checking algorithm for CTL but specifically
target parallel, shared memory machines. Our two versions differ by
the amount of information they need to store: a Reverse Graph (RG)
version that explicitly stores the complete transition relation in
memory, and a Reverse Parental Graph (RPG) that relies on a spanning
tree.

We use the reverse parental graph structure as a mean to fight the
state explosion problem. In this respect, this approach has a similar
impact---on the space---than algorithmic techniques like \emph{sleep
  sets} (used with partial-order methods), but with the difference
that we do not take into account the structure of the model. Moreover,
our approach is effective regardless of the formalism used to model
the system. For instance, it is particularly useful in cases where it
is not possible to compute the ``inverse'' of the transition relation.

Our prototype implementation shows promising results for both the RG
and RPG versions of the algorithm. The choice of a ``labeling
algorithm'' based on the out-degree number has proved to be a good
match for shared memory machines and a work stealing strategy; 
we consistently obtained speedups close to linear with an
average efficiency of $75\%$.  Our experimental results also showed
that the RPG version is able to outperform the RG version for some
categories of models.

Using our work, one can easily obtain a parallel algorithm for checking
any CTL formula $\Phi$ by running one instance of our algorithms (for
the AU and EU formulas) for each subformula of $\Phi$. But this
approach, as such, is too naive. For future works, we are considering
improvements of our algorithms that support full CTL formulas without
having to manage several copies of our labels ($\sons$ and $\suc$) in
parallel, which could have an adverse effect on memory consumption.

\begin{small}
\bibliographystyle{plain}
\bibliography{ref}
\end{small}


\end{document}